\documentclass[aps,prc,preprint,showpacs]{revtex4-1}
\usepackage[dvips]{graphicx}
\usepackage{dcolumn}
\usepackage{bm}

\begin{document}

\def\JAGU{M. Smoluchowski Institute of Physics, Jagiellonian University, Pl-30-348 Krak\'ow, Poland}
\def\GSI{GSI Helmholtzzentrum f\"{u}r Schwerionenforschung GmbH, D-64291 Darmstadt, Germany}

\title{Distributions of the largest fragment size in multifragmentation: 
Traces of a phase transition}

\author{J.~Brzychczyk}     \email{janusz.brzychczyk@uj.edu.pl}      \affiliation{\JAGU}   
\author{T.~Pietrzak}            \affiliation{\JAGU}   
\author{W.~Trautmann}           \affiliation{\GSI}
\author{A.~Wieloch}             \affiliation{\JAGU}

\date{\today}

\begin{abstract}
Distributions of the largest fragment charge are studied
using the ALADIN data on fragmentation of $^{197}$Au projectiles
at relativistic energies.
The statistical measures skewness and kurtosis of higher-order fluctuations provide a robust indication of
the transition point, linked to a phase transition in the thermodynamic limit.
Extensive comparisons with predictions of a bond percolation model
corroborate the high accuracy of this model
in reproducing the distributions as well as the whole fragmentation pattern
as represented by the measured charge correlations.
In analogy to percolation, the pseudocritical and critical points are identified in the fragmentation data.
Questions concerning the distinction between different models
and between first- and second-order phase transitions are discussed.
\end{abstract}


\maketitle

%\newpage

\section{Introduction}
In nuclear multifragmentation studies, the mass of the largest fragment and its distribution have received special attention.
The largest fragment is often identified with the liquid phase in a mixed-phase configuration and thus assumed to play 
the role of the order parameter in a liquid-gas phase-transition scenario. Its distributions are expected to provide 
valuable insight into the phase behavior of the investigated 
systems~\cite{d-p,uf,carmona,zhang04,frankland,ma05,gulm2005,pichon06,leneindre2007,bonnet2009}.
A transition from a ``liquid'' to a ``gaseous'' state is associated with a rapid decrease of the largest fragment size. 
It may correspond to the order-parameter discontinuity in the case of a first-order phase transition or to the power-law
disappearance near a second-order transition point (a critical point).
Besides the characteristic evolution of its mean-value, event-to-event fluctuations reflected in the probability distribution
of the largest fragment size are of considerable interest. Experimental examinations have focused on the appearance of 
particularly large fluctuations~\cite{c2,brzy98,dorso,ell,hyp,ma05,leneindre2007},
on bimodal characteristics representative of a two-phase coexistence~\cite{bord,ma05,pichon06,lopez,chaud07,bonnet2009},
on so-called $\Delta$-scaling features~\cite{d-p,frankland,ma05,pichon06},
and on the connection with dynamical observables as, e.g., radial flow~\cite{gruyer13}.

Although the presence of a phase transition is often deduced, its kind is usually not unambiguously identified.
In small systems the asymptotic behavior is strongly modified by finite size and surface effects, so that
the distinction between first- and second-order phase transitions becomes very difficult.
Simulations with lattice gas models have shown that critical-like features are observed in finite systems not only 
along the Kert\'{e}sz line~\cite{kertesz,campi1997} but also inside the liquid-gas coexistence region, 
i.e. the first order phase transition can mimic critical behavior~\cite{carmona,gulm2005,mim,das2002,wieloch}.  
Moreover, the control parameter, the temperature or energy content in a thermodynamical phase transition,
cannot be precisely measured and must be substituted by a another measurable quantity. Sorting events according
to a substitute control paprameter will additionally blur the observed signals.

The present work is motivated by percolation studies suggesting new signatures of a critical behavior associated with
the distribution properties of the largest fragment size or mass~\cite{jb}. They are exhibited by
the cumulant ratios up to fourth order, i.e. the normalized variance, skewness and kurtosis.
Specific features of these dimensionless cumulants characterizing the distribution provide a robust indication of the 
pseudocritical point in finite systems and permit estimates of the location of the critical point in the continuous limit.
Fluctuation observables up to fourth order have also been proposed for probing the QCD chiral transition and for
searching for the QCD critical point with experimental data at much higher energies~\cite{steph11,redlich}. They
are presently widely used to characterize
the hot medium generated in ultra-relativistic heavy-ion collisions~\cite{asakawa,feckova}. The rich phenomenology of
fluctuation observables near a critical point has recently been explored in a description of the liquid-gas phase
transition of nuclear matter with a model based on the Van-der-Waals equation~\cite{vov2015}. 

These new signatures are applicable to very small systems and can be tested with various measurable sorting
variables. Therefore, they are well suited for an application to nuclear multifragmentation.
Within this context, properties of the largest fragment size
in multifragmentation are investigated using the results of the ALADIN Collaboration~\cite{sch}.
In addition, comparisons between predictions of a bond percolation model and the experimental data are not restricted
to the largest fragment characteristics~\cite{kreutz}. The whole fragmentation pattern is verified in detail to obtain 
a quantitative reference, permitting comparisons with other models representing alternative multifragmentation scenarios.

A brief recall of the main cumulant features near the percolation transition is given in the next section.
To obtain some information on the question of their uniqueness or universality, percolation results are compared 
with predictions of a thermodynamic model known to contain a first-order phase transition~\cite{the,gupta98,chaud07} .

\section{Distribution of the largest fragment size}

Percolation calculations presented in \cite{jb} and in
this work are performed
with a three-dimensional bond percolation model on simple cubic
lattices \cite{sta,b1}.
Events are generated using a Monte Carlo procedure.
The sites are arranged on the lattice
in the most compact configuration, the bonds are created randomly
with probability $p$. Clusters are recognized with 
the Hoshen-Kopelman algorithm \cite{h-k}.
Free boundary conditions are applied to account for
the presence of surface in real systems.

Given a control parameter value $p$ and 
the total number of sites $A_{0}$ (the system size),
the probability distribution $P(A_{max})$
of the largest cluster size $A_{max}$ is determined from
a large sample of events.
The statistical measures as the mean, variance,
skewness and kurtosis contain the most significant
information about the distribution.
Of particular interest are the following dimensionless
cumulant ratios
\begin{eqnarray}
K_2\equiv &\mu_{2}/\langle A_{\rm max}\rangle^2& =\kappa_2/\kappa_1^2 \nonumber\\
K_3\equiv &\mu_{3}/\mu_{2}^{3/2}& =\kappa_3/\kappa_2^{3/2} \nonumber\\
K_4\equiv &\mu_{4}/\mu_{2}^{2}-3& =\kappa_4/\kappa_2^{2},
\end{eqnarray}
where $\langle A_{max}\rangle$ denotes the mean value,
$\mu_i=\langle(A_{max}-\langle A_{max}\rangle)^i\rangle$ is the $i$th
central moment, and $\kappa_i$ is the $i$th cumulant of $P(A_{max})$.
$K_{2}$ is the variance normalized to the squared mean, 
$K_{3}$ is the skewness which indicates the distribution asymmetry, and
$K_{4}$ is the kurtosis excess measuring the degree of peakedness.
The cumulants are simple functions of the central moments with
$\kappa_1 = \langle A_{max}\rangle$, $\kappa_2 = \mu_2$, $\kappa_3 = \mu_3$,
and $\kappa_4 = \mu_4 - 3\mu_2^2$.
In the transition region, these quantities obey with good accuracy
finite-size scaling relations even for very small systems
with open boundaries~\cite{jb}. This permits the identification of universal
(independent of the system size) features of $K_{i}$ at the transition point or region.

\begin{figure}[ht]
\includegraphics[width=12.3cm]{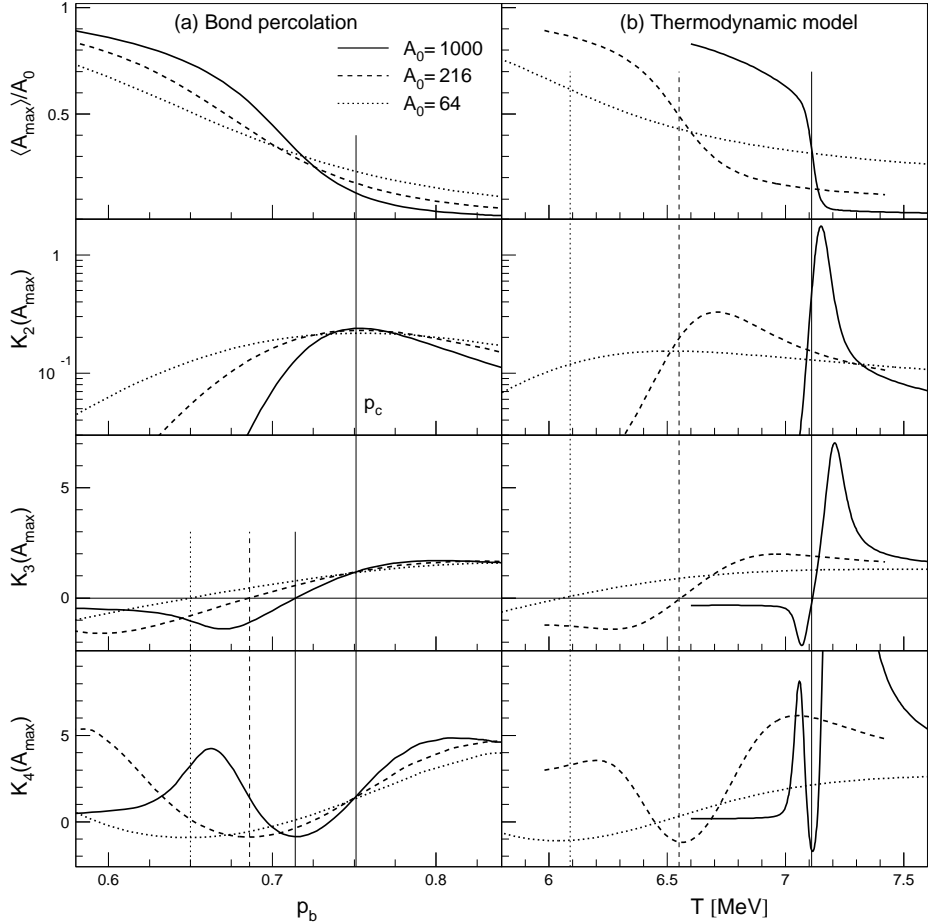}
\caption{The cumulants of Eqs. (1) for three different system sizes.
(a) Percolation results plotted as a function of the bond breaking
probability. The long vertical line indicates the critical point $p_{c}$
in the continuous limit. The short lines indicate
the transition (pseudocritical) points for the finite
systems. (b) Results of the thermodynamic model as a function of
the temperature. The vertical lines mark the transition temperatures
corresponding to the maximum specific heats of the three systems. 
For $A_{0} = 1000$, the kurtosis excess $K_{4}$ reaches a maximum value 78 at $T \simeq 7.2$~MeV (bottom right panel).}
\label{fig:1}
\end{figure}

This form of universality can be illustrated with the help of Fig.~1(a).
The cumulant ratios $K_{i}$ are
plotted as a function of the bond breaking probability
$p_{b}\equiv 1-p$ for three different system sizes.
Increasing $p_{b}$ corresponds to increasing the temperature in a physical
application that contains temperature as a control parameter, as it is the
case here for nuclear multifragmentation.
The location of the critical point in the continuous limit
$p_{c}\simeq 0.751$ is marked by the vertical long line.
According to finite-size scaling, the values of the cumulants $K_{i}$ 
at $p_{c}$ are expected to be independent of the system size.
This is quite precisely observed in Fig.~1(a) as the crossing of the curves.
A prominent feature of $K_2$ is its maximum located very close to $p_c$.
Maxima of other quantities used as criticality signals show much larger deviations from $p_c$ 
(e.g., the maximum variance of the fragment mass distribution, see examples given in Ref.~\cite{jb}). 
The transition point in finite systems can be associated with the broadest and 
most symmetric $P(A_{max})$ distribution
observed near the pseudocritical point defined by the maximum of
the mean cluster size being the analog of the susceptibility~\cite{jb}.
This transitional distribution is indicated by
$K_{3}=0$ and the minimum value of $K_{4}$ of about $-1$.
Figure 2(a) shows examples of such distributions.
The distance of the pseudocritical point from $p_c$ increases with
decreasing system size according to finite-size scaling (Fig.~1(a)).

\begin{figure}[ht]
\includegraphics[width=8.5cm]{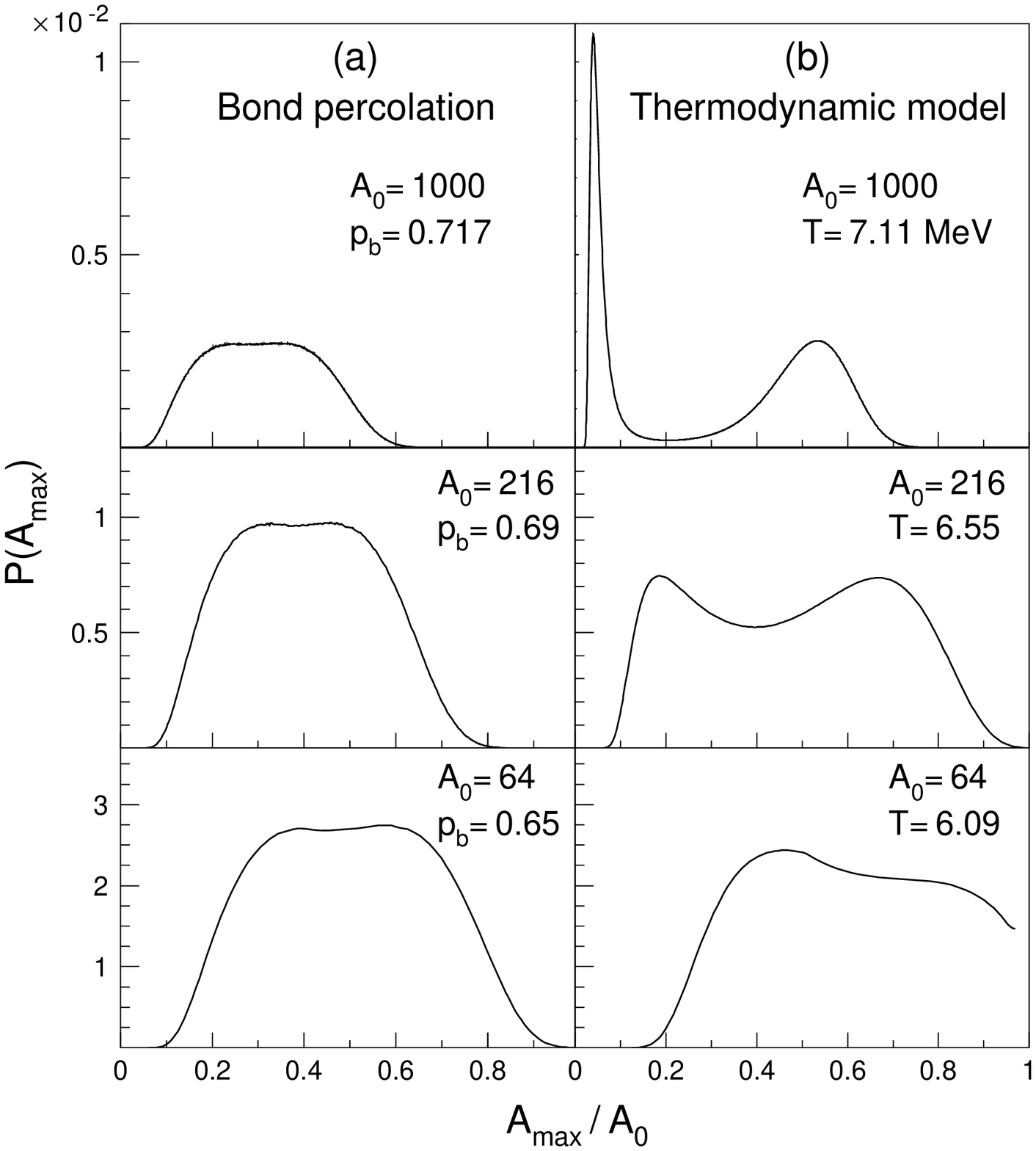}
\caption{Probability distributions of the largest fragment size
at the transition points.}
\label{fig:2}
\end{figure}
The cumulant features characterizing the critical and
pseudocritical points are approximately
preserved for the corresponding points when events are sorted
by measurable variables correlated with
the control parameter, such as the total multiplicity or
the total mass of complex fragments.
Near-critical events are indicated by the maximum of $K_2$
while events associated with the pseudocritical point are characterized
by $K_{3}=0$ and the minimum value of $K_{4}$ of about $-1$.

It is instructive to compare the percolation results
with predictions of the thermodynamic model~\cite{gupta98} 
which is a simplified version of the statistical
multifragmentation model (SMM, Ref.~\cite{smm}).
The calculations have been performed for the canonical ensemble
of noninteracting one-component fragments.
The model permits computing the partition function,
and thus to obtain the thermodynamic properties of the system.
The presence of a first-order phase transition
is well established \cite{the,gupta98,bug}.
The results for a freeze-out density of one third of the
normal nuclear density are presented in Fig.~1(b).
The transition temperatures for
systems with $64$, $216$, and $1000$ nucleons, derived
from the locations of the specific heat maximum, 
are $6.09$, $6.55$, and $7.11$ MeV, respectively.
These values are marked by the vertical lines.
In the thermodynamic limit, the maximum location is expected at $T \simeq 8$~MeV
as calculated within the grand canonical approach 
and shown in Fig.~4 of Ref.~\cite{bug}.
In this model, the critical temperature is assumed to be $T_c=18$~MeV.

As can be seen in the top diagram of Fig.~1(b), $\langle A_{\rm max}\rangle$
shows the fastest decrease at the transition point.
A step discontinuity develops with increasing system size~\cite{gupta98}.
Similarly to the case of percolation, the transition point is precisely indicated by
$K_{3}=0$ and the minimum of $K_{4}$. Here, $K_{4}$ reaches somewhat lower values, 
suggesting a bimodal structure of the probability distribution
as expected for a first-order phase transition in the canonical ensemble. 
Figure~2(b) shows that, for a system as large as $A_{0}=1000$, a distinct bimodality is observed.
The two-peak structure gradually vanishes as the system size decreases.

The above examples illustrate that the transition point in small systems, 
associated with a phase transition in the thermodynamic limit, is well indicated
by $K_{3}=0$ together with a minimum of $K_{4}$ for both, i.e. first- and second-order, types of transition.
The comparisons suggest that some evidence for
the transition order can be obtained from the evolution
of $P(A_{max})$ with the system size.
In particular, in the vicinity of the transition point
the cumulants exhibit maxima whose amplitudes 
increase with the system size
in the case of the thermodynamic model.
This is in contrast to percolation where the amplitudes are bounded
according to the second-order finite-size scaling.

\section{Experimental data}

The present work examines the ALADIN data on fragmentation of projectile
spectators in $^{197}$Au + Cu, In, Au peripheral collisions at
the incident energies of 600{\it{A}}~MeV (Cu, In, Au targets),
800{\it{A}}~MeV (Au), and 1000{\it{A}}~MeV (Cu, Au).
Details of the experiment and general characteristics of the data have been presented in Ref.~\cite{sch}.
Fragments with the atomic numbers $Z\geq2$ were detected with high
efficiencies, close to 100\% at the bombarding energy of 1000 MeV/nucleon, and fully $Z$-identified. 
For the present work, the event-sorted data files were used that formed the basis of the results reported in
the experimental paper~\cite{sch}. The magnitude of potential effects caused by the minute but finite acceptance 
losses on higher-order correlations were investigated and results will be shown below.

\begin{figure}[ht]
\includegraphics[width=8.8cm]{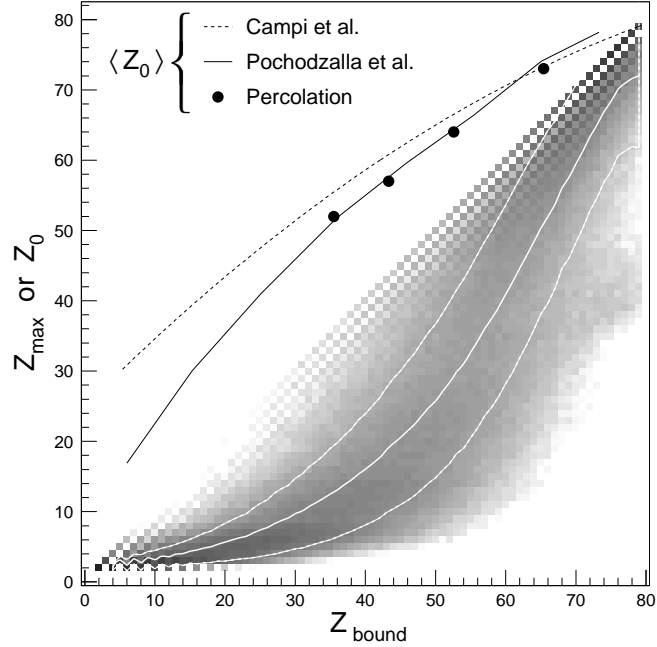}
\caption{Distribution of $Z_{\rm max}$ vs $Z_{\rm bound}$
for the ALADIN data (with shadings in a logarithmic scale).
The solid white line shows the mean values of $Z_{\rm max}$,
the broken lines indicate the rms dispersions.
The full and dashed black lines represent two estimates of the mean system size $Z_0$ 
obtained from the experimental data with different 
methods~\protect\cite{poch,campi94} while the filled circles give the result
obtained with percolation (see text).} 
\label{fig:3}
\end{figure}

It is a prominent feature of the data that the fragment multiplicities and correlations are independent of
the projectile energy and the target nucleus when plotted
as a function of $Z_{\rm bound}$. This universality has been interpreted as indicating a high degree
of equilibration attained prior to or during the fragmentation stage~\cite{sch}.
The quantity $Z_{\rm bound}$ is defined
as the sum of the atomic numbers $Z$ of all fragments with $Z\geq2$.
It serves as a sorting variable, correlated with the size of the projectile
spectator and inversely correlated with the excitation energy per nucleon~\cite{hub}. 

The correlation between the largest fragment charge, $Z_{\rm max}$,
and $Z_{\rm bound}$ is illustrated in Fig.~3. All the studied data sets are included.
At low excitation energies, corresponding to largest $Z_{\rm bound}$
values, evaporation processes are dominant (events with one large fragment).
There is also a small fraction of fission events with $Z_{\rm max}$ around $40$.
At excitation energies approaching and exceeding the nuclear binding energy
($Z_{\rm bound}<40$), the systems are disassembled into many
small fragments. The transition between the two extreme regimes
is characterized by a rapidly decreasing $Z_{\rm max}$ associated with an
increasing number of fragments.

Experimental information on the size of the fragmenting system
(spectator remnant) is important for testing theoretical predictions.
An estimation of the mean system mass
$\langle A_{0}\rangle$ as a function of $Z_{\rm bound}$ in several
$Z_{\rm max}$ windows
was made for the 600{\it{A}} MeV $^{197}$Au + Cu reaction \cite{poch}.
Assuming the charge-to-mass ratio $Z_{0}/A_{0}=0.4$,
the value of $^{197}$Au projectiles, the results
converted to the mean system charge $\langle Z_{0}\rangle$ and averaged over
$Z_{\rm max}$ bins are shown in Fig.~3 by the solid line.
A similar result (dashed line) was obtained by Campi {\it et al.} who used
a sum-rule approach for extrapolating from the measured $Z_{\rm bound}$ 
to $Z_{0}$~\cite{campi94}.

In the present work the system sizes will be deduced from
comparisons between the experimental data and the predictions
of the percolation model. These results are
indicated in the figure by the filled circles.

\section{Percolation analysis}

In the experimental data, the atomic number serves as a measure
of the fragment size. 
The corresponding measure in the percolation analysis performed here
is the number of sites contained within a cluster. 
In other words, the number of sites is considered as the number of proton charges.
In the following,
the same notation will be used for percolation quantities
as for their experimental counterparts.
Percolation events are generated for the bond probabilities
uniformly distributed in the interval $[0,1]$, and then sorted
according to $Z_{\rm bound}$.

The cumulant ratios $K_{i}$ of the largest-fragment size distribution $P(Z_{\rm max})$
are examined in Fig.~4. The percolation results are plotted
in the left part as a function of $Z_{\rm bound}$, normalized to the
system size $Z_{0}$, for three different system sizes that span
over a range expected to be in the transition region.
In this representation, the $K_{i}$ distributions show a weak dependence on the
system size which vanishes at the pseudocritical point located
at $Z_{\rm bound}/Z_{0}\simeq 0.84$ (the squares).
The results corresponding to the true critical point are located near the maximum
of $K_{2}$ (cf. Fig.~1) but their positions on the $Z_{\rm bound}/Z_{0}$ axis depend somewhat on 
the system size (filled circles in Fig.~4, left panel).

\begin{figure*}[ht]
\includegraphics[width=14.2cm]{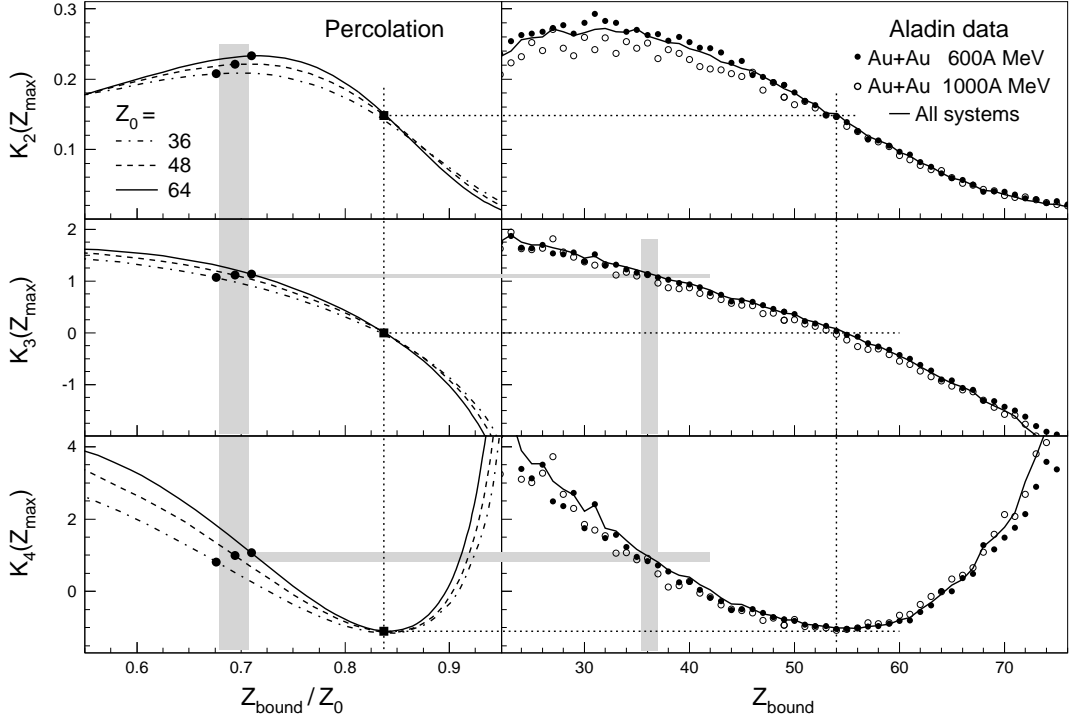}
\caption{The cumulants of $P(Z_{\rm max})$ as a function of $Z_{\rm bound}$ (or $Z_{\rm bound}$
normalized with respect to $Z_0$ in the case of percolation). The results of
bond percolation calculations for three values of $Z_0$ (left panels) are shown in comparison with the 
experimental data for $^{197}$Au fragmentation (right panels) following collisions with $^{197}$Au targets
at two energies (symbols) and the combined results for all systems (full lines).
The meaning of the lines and of the symbols in the left panels is explained in the text. 
The statistical errors are of the order of the scatter of the data symbols (see text).}
\label{fig:4}
\end{figure*}

The experimental results are shown in the right diagrams for
the $^{197}$Au + $^{197}$Au systems at 600 and 1000 MeV/nucleon
and for the summed data sets (all targets and all energies).
Here, the cumulant ratios $K_{i}$ are plotted as a function of $Z_{\rm bound}$.
The system sizes are considered unknown quantities that are to be determined.
The comparison of the different data sets indicates significant
systematic differences only for $K_{2}$ below $Z_{\rm bound}\simeq 50$.
The statistical errors are small and comparable to the apparent scatter of the
data points. They are smallest near the pseudocritical point $Z_{\rm bound} = 54$ and there smaller
than the size of the data symbols. Larger errors are expected for smaller $Z_{\rm bound}$. At $Z_{\rm bound} = 31$,
e.g., the error analysis for the $^{197}$Au + $^{197}$Au system at 1000 MeV/nucleon 
yields 0.0091, 0.077, and 0.347 for the statistical uncertainties of $K_{2}$, $K_{3}$, and $K_{4}$, respectively. 
For clarity, these errors are not displayed in the figure.

\begin{figure*}[ht]
\includegraphics[width=8.5cm]{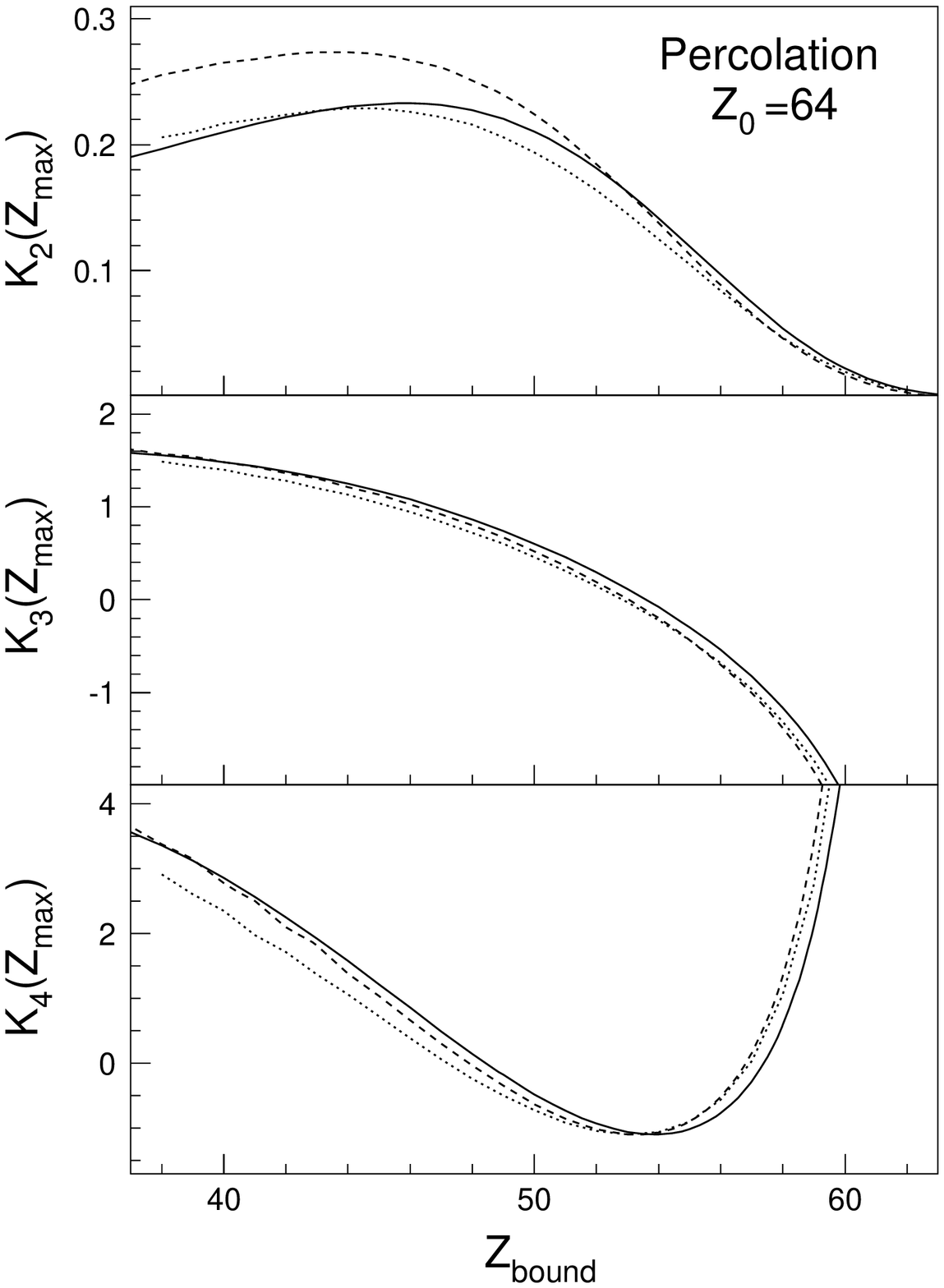}
\caption{The cumulants of $P(Z_{\rm max})$ as a function of $Z_{\rm bound}$ 
calculated for the system size $Z_0 = 64$ as shown in the left panel of Fig. 4 (full lines)
in comparison with the results of the two tests described in the text investigating the effects of 
secondary decays (dashed lines) and of the detection efficiency of the spectrometer (dotted lines).}
\label{fig:5}
\end{figure*}

Overall, the percolation and experimental patterns of $K_{i}$ are very similar.
The specific characteristics of the percolation pseudocritical point
are well observed in the data at $Z_{\rm bound}\simeq 54$.
With this correspondence, the mean system size at $Z_{\rm bound}\simeq 54$
may be estimated as $Z_{0}\simeq Z_{\rm bound}/0.84\simeq 64$.
For the percolation ``critical'' point, an approximate
correspondence can be established relying on $K_{3}$ and $K_{4}$.
It indicates $Z_{\rm bound}\simeq 36$ and a system size $Z_{0}$ around $36/0.69 \simeq 52$.

The experimental values of $K_{2}$ in the region of small $Z_{\rm bound}$
depend slightly on the projectile energy. It may be related to a sensitivity of $K_{2}$ 
to existing small changes in the reaction dynamics or perhaps simply
to a residual energy dependence of the detection efficiency. 
At small $Z_{\rm bound}$, i.e. at large excitation energies, secondary evaporation
effects may also be of importance.
In order to test the sensitivity of $K_{i}$ to such effects a simple simulation was performed and 
applied to percolation events. The result is shown in Fig.~5. The largest cluster was divided
into two fragments: $Z_{\rm max}\rightarrow (Z_{max}-1)+1$ with
probability $p_{1}$, or $Z_{max}\rightarrow (Z_{max}-2)+2$ with
probability $p_{2}$. Such calculations with various assumptions
on $p_{1}$ and $p_{2}$ indicate that $K_{2}$ significantly
increases at small $Z_{\rm bound}$ while $K_{3}$ and $K_{4}$
remain nearly unchanged. As an example, for $Z_{0}=64$ with $p_{1}=0.6$
and $p_{2}=0.3$, the maximum of $K_{2}$ increases from 
$0.23$ to $0.27$ and its position is shifted towards lower
$Z_{\rm bound}$ by $\sim 5\%$ (Fig.~5, dashed lines).
These observations support the conclusion that $K_{3}$ and $K_{4}$ are
more reliable than $K_{2}$ as quantitative indicators of the transition points.

\begin{figure}[ht]
\includegraphics[width=8.2cm]{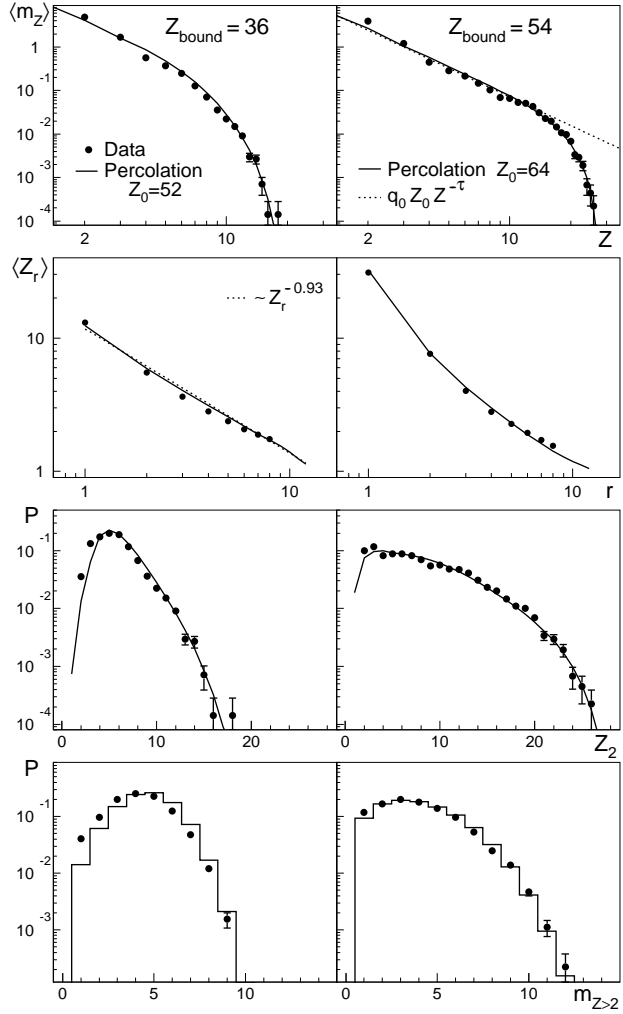}
\caption{Comparison of percolation predictions (full lines) for $Z_{0}=52$ (left panels)
and $Z_{0}=64$ (right panels), with the experimental data for all systems (filled circles), selected 
with the conditions $Z_{\rm bound}=36$ and $54$, respectively.
From top to bottom: the mean fragment multiplicity as a
function of the fragment size $Z$ (the largest fragment excluded),
the mean fragment size as a function of the fragment rank $r$,
the probability distribution of the size of the second largest
fragment $Z_2$, and the multiplicity distribution $m_{Z>2}$ of fragments with $Z>2$.
The dotted lines represent power-law descriptions as indicated.
The statistical errors are shown where they are larger than the data symbols.}
\label{fig:6}
\end{figure}

The results of a test investigating the effects of the small detection inefficiencies of the spectrometer
are shown in the same panels of Fig.~5 (dotted lines). It consisted of modifying the percolation event files
by randomly deleting fragments with probabilities $1-\epsilon (Z)$ and redoing the cumulant analysis with
these modified files. The detection efficiencies $\epsilon (Z)$ were assumed to have the values of the
geometrical acceptance of the time-of-flight wall of the spectrometer as determined for 800~MeV/nucleon
incident energy~\cite{sch-thesis}. They increase smoothly from $\epsilon (2) = 0.93$ to 
$\epsilon (7) = 0.99$ and $\epsilon (Z) = 1.00$ for $Z \ge 8$. This test cannot restore the original
event structure nor can it take account of additional sources of uncertainties mentioned
in the experimental reports~\cite{sch,sch-thesis} as, e.g., reactions in the detector material. However, these processes are
estimated to cause similarly small effects on the percent level whose magnitude can equally be estimated
from the deviation of the test result. Both tests indicate that the modifications can be expected to be small, 
corresponding to an uncertainty of the order of one unit of $Z_{\rm bound}$ on the abscissa.
The coincidence of the zero crossing of $K_3$ with the minimum of $K_4$ is not affected.

The pairwise correspondence exhibited by the $K_{3}$ and $K_{4}$
cumulant ratios implies very similar $P(Z_{max})$ distributions but is even 
more general.
A close resemblance between the whole fragmentation patterns
is observed, as demonstrated in Fig.~6 for the data sets with $Z_{\rm bound}=36$ and $54$. 
The percolation calculations are performed for the estimated
system sizes $Z_{0}=52$ and $64$, respectively.
All the experimental data sets are summed for better statistics.

The top row of panels shows the fragment size distributions.
The model describes the data rather well over four orders
of magnitude. At $Z_{\rm bound}=54$, as expected for the percolation
pseudocritical point, the distribution follows for $Z<15$
the asymptotic power-law dependence with the exponent
$\tau=2.189$ and the normalization constant $q_{0}=0.173$~\cite{jb}.

The next row shows the Zipf-type plots, i.e. the mean size
of the largest, second largest, and up to the $r^{th}$-largest fragments plotted
against their rank $r$.
Such plots have been examined in the context of the expected appearance of
Zipf's law near a critical point \cite{ma-z,ma05,dabr,campi-z,ba-p,leneindre2007}.
The Zipf's law states that $\langle Z_{r}\rangle\sim 1/r^{\lambda}$
with $\lambda\simeq 1$.
The percolation results are shown for the rank numbers $1$ to $11$. 
The total fragment multiplicity $m$ is at least $11$ in all the percolation events.
This is implied by the condition $m>Z_{0}-Z_{\rm bound}$.
The experimental data contain information only on fragments with $Z>1$.
Since their mean multiplicities for $Z_{\rm bound}=36$ and $54$
are about $9.1$ and $7.7$, the mean total multiplicities including $Z=1$ isotopes
may be estimated as $25.1$ and $17.7$, respectively, by assuming $m_{Z=1} = Z_{0}-Z_{\rm bound}$.    
It was, therefore, assumed
that events containing less than 8 fragments can be supplemented
with fragments of $Z=1$ up to the rank of 8.
At $Z_{\rm bound}=36$, one observes an approximate behavior according to Zipf's
law  with the exponent $\lambda\simeq 0.93$
determined from a fit to the percolation results.
It is worthwhile to note that this feature appears
at the ``critical'' point while a trace of the asymptotic power-law behavior 
in the fragment size distribution (with the largest fragment excluded)
is observed at the pseudocritical point $Z_{\rm bound}=54$. 

The percolation model very well describes not only mean values
but also event-to-event fluctuations.
As an example, the third row of panels of Fig.~6 shows the probability (yield fraction)
distributions of the second largest fragment.
The bottom row shows the multiplicity distributions of fragments with $Z>2$.
At $Z_{\rm bound}=36$ where the excitation
energy is large, the calculated multiplicity distribution is shifted to values
slightly larger than the experimental results while they practically coincide at $Z_{\rm bound}=54$. 

\begin{figure}[ht]
\includegraphics[width=9.5cm]{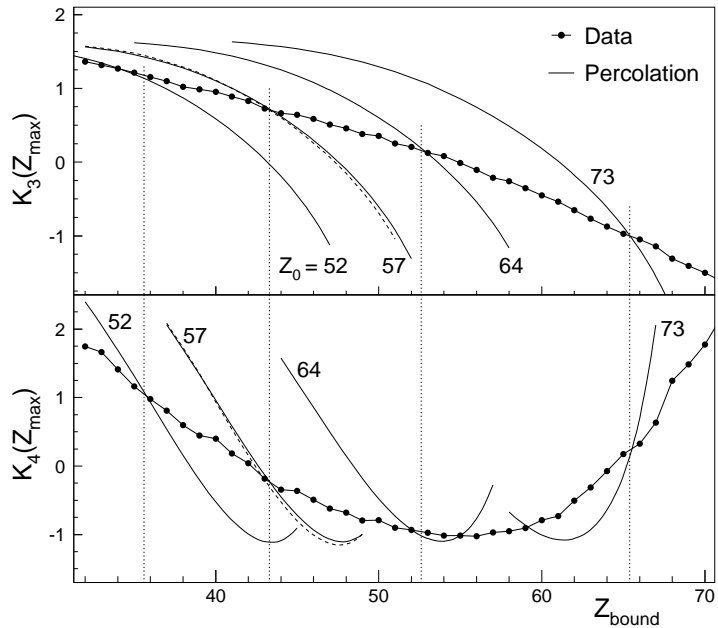}
\caption{The $K_{3}$ and $K_{4}$ cumulants as a function of 
$Z_{\rm bound}$. The experimental data (filled circles, all systems, as shown by full lines in Fig. 4, right panel) and
percolation calculations for the indicated system sizes $Z_0$. 
The dotted vertical lines illustrate the identical crossing points
for $K_{3}$ and $K_{4}$, the dashed line shows the result obtained with a Gaussian distribution 
of $Z_0$ with $\langle Z_0 \rangle = 57$ (see text).}
\label{fig:7}
\end{figure}

For such comparisons that can be extended to other $Z_{\rm bound}$
values, only the knowledge of the relation between $Z_{\rm bound}$ and $Z_{0}$ is required.
Here, this relation is determined on the basis of the $K_{3}$ and $K_{4}$ equalities. 
The analysis can also be performed in a slightly different manner.
In Fig.~7 the cumulants are plotted as a function of $Z_{\rm bound}$ for the experimental data and
percolation systems of different sizes.
For a given $Z_{0}$, the crossing of the percolation and experimental lines determine the corresponding $Z_{\rm bound}$.
The relation found by this procedure is displayed in Fig.~3 by the solid circles.
It shows good agreement with the experimental estimates of Ref.~\cite{poch}.

In reality, the system sizes at fixed $Z_{\rm bound}$ are dispersed.
To evaluate the significance of this dispersion for the analysis, 
percolation calculations have been performed for
a Gaussian distribution of $Z_{0}$ with the mean of $57$ and
standard deviation of $2$. Such a deviation is suggested
by SMM simulations performed with input conditions established
in Ref.~\cite{botv}. The results are plotted in Fig.~7 by the dashed lines, 
showing that the dispersion effects are not substantial.

\begin{figure*}[ht]
\includegraphics[width=12.cm]{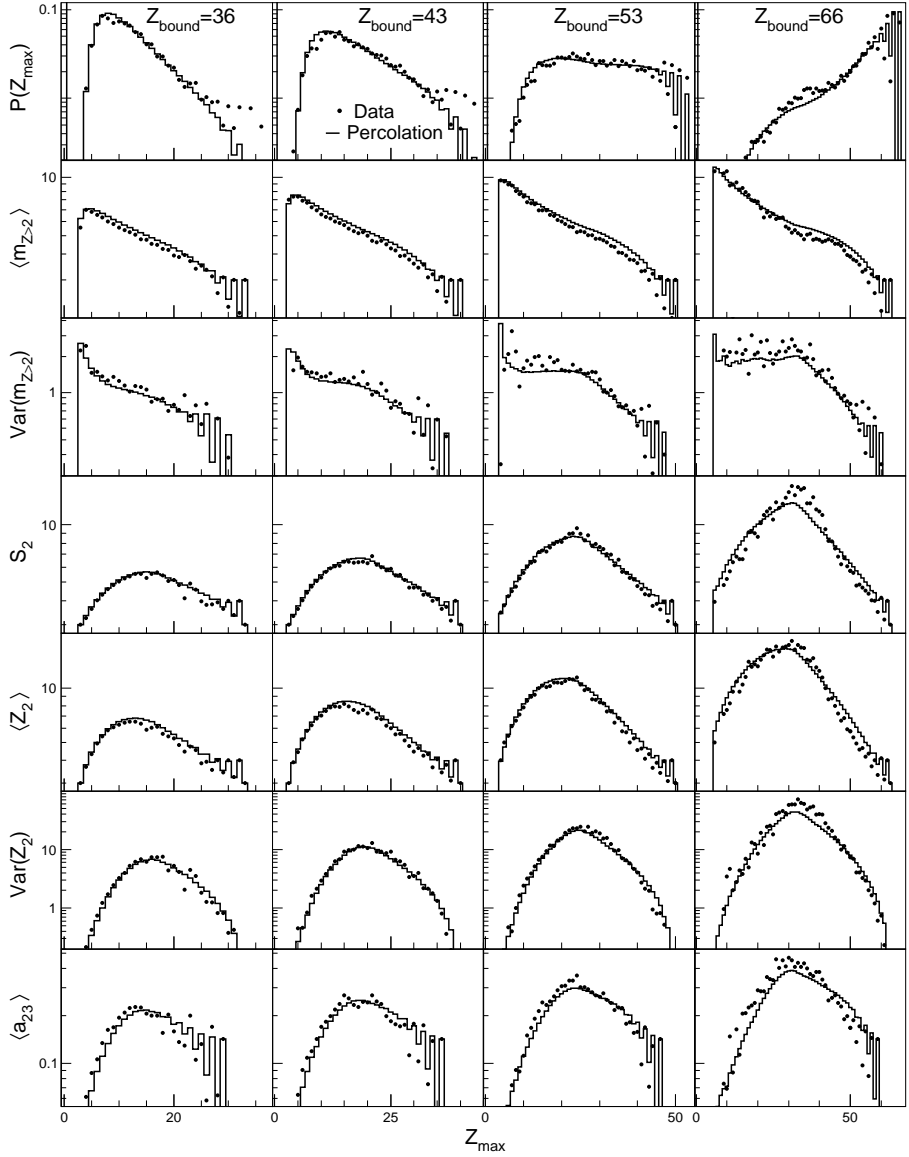}
\caption{Percolation (solid histograms) vs experimental data (dots, all systems):
characteristics of fragments with $Z>2$ are
plotted as a function of $Z_{max}$.
From top to bottom: the $Z_{max}$ distribution,
the mean and variance of the fragment multiplicity,
the mean fragment size, the mean and variance of
the second largest fragment size, the mean size
asymmetry between the second and third largest
fragments. The statistical errors are of the order of the scatter of the data symbols.}
\label{fig:8}
\end{figure*}

The model predictions are examined in more detail in Fig.~8.
The comparisons are made at the matching conditions determined from Fig.~7.
To not obscure the histograms, errors bars are omitted in the figure.
Their magnitude is evident from the scatter of the data symbols.
The top diagrams show the $P(Z_{\rm max})$ distributions.
Overall, the agreement is very good, although the experimental
distributions exhibit some local enhancements which are not
accounted for by the model. 
They are seen for $Z_{\rm bound}=36$ and $43$ at largest $Z_{\rm max}$,
which corresponds to events with one large fragment and few
light particles. Such evaporation-like events are not unexpected
even for large excitation energies, since neutron-rich projectile
spectators can be cooled by neutron emissions.
Another enhancement which is seen at $Z_{\rm bound}=66$ near
$Z_{\rm max}\simeq 36$ is most likely associated with a contribution from fission events.

The next panels in Fig.~8 examine various fragment size
characteristics and correlations as a function of $Z_{\rm max}$.
They are the mean and variance of the multiplicity of fragments
with $Z>2$, the mean fragment size $S_{2}$ defined as
the second moment of the fragment size distribution normalized
to the first moment (the largest fragment excluded) \cite{sta},
the mean and variance of the second largest fragment size,
and the size asymmetry between the second and third largest
fragments $a_{23}\equiv(Z_{2}-Z_{3})/(Z_{2}+Z_{3})$ averaged over events
with $Z_{3}>2$.
In these calculations only fragments with $Z>2$ are taken
into account to avoid contributions from light particles coming
from other sources.

One may ask whether simultaneously fixed $Z_{\rm bound}$ and $Z_{\rm max}$
will severely limit the possibilities for the fragment-size partitioning,
leading to rather trivial results.
This is the case when $Z_{\rm max}$ is close to its limiting
minimum or maximum value.
The number of possible partitions is largest for $Z_{\rm max}$
in the middle of its range, around $Z_{\rm bound}/2$. 
This least restrictive condition allows for a better test
of fragmentation patterns.
Without additional $Z_{\rm max}$ selection the characteristics
would be dominated by the trivial contributions in the cases when
$P(Z_{\rm max})$ distributions are peaked near the limiting $Z_{max}$ values
(cf. top rows of Fig.~8).

This very detailed quantitative analysis confirms that
the bond percolation model remarkably well describes the experimental fragment
sizes and their fluctuations. This has been noticed early on after the first
ALADIN data on projectile fragmentation had become available and was reported
for a set of fragment distributions and asymmetries in Ref.~\cite{kreutz}.
The percolation results may serve as a reference
for further analysis with other models and for sensitive tests of their performance. 
In fact, calculations performed with the
lattice-gas model and with the SMM have shown that the characteristic coincidence
of the zero transition of the skewness and the minimum of the kurtosis excess
is observed with these models as well~\cite{wieloch, pietrzak}. Considered as
indicators for a second-order phase transition, they are simultaneously present
in these statistical models that are believed to exhibit a phase
transition of first-order in the thermodynamic limit.

The cumulant analysis indicates $Z_{\rm bound}=54$ as the transition point for nuclear systems with $Z_{0}\simeq 64$.
Based on estimations performed for the $^{197}$Au + $^{197}$Au reaction at $600$ MeV/nucleon
\cite{sch,poch}, this point corresponds to the excitation energy
around $6$ MeV per nucleon, which can be associated with a temperature
within the range $5-7$~MeV~\cite{poch,nat,temp}.                         
In the percolation context, the transition point can be interpreted as the pseudocritical point.
The analysis suggests that near-critical events are rather located
at $Z_{\rm bound}\simeq 36$ for which the estimated excitation energy is
about $10$~MeV per nucleon, corresponding to temperatures in the range 6 to 7~MeV. 
These temperatures are much lower than the critical temperature of about $14-15$~ MeV,
calculated with relativistic mean-field models for
asymmetric nuclear matter with a proton fraction of the $^{197}$Au nucleus~\cite{horst,guang}.
However, in finite nuclear systems the critical temperature can be reduced by more than 5~MeV
due to the presence of the Coulomb and surface effects~\cite{jaqaman}.
The estimated temperatures depend on the method and the size of the studied system.
For example, according to calculations with the Fermionic Molecular Dynamics model performed for $^{16}$O,
the critical temperature deduced from observing the disappearance of the liquid-gas coexistence
is about $10$~MeV~\cite{schnack}. A somewhat larger value $T_{c} \simeq 12$~MeV has been concluded from a study
of a system of mass number $A = 36$ with antisymmetrized molecular dynamics~\cite{furuta}.

\section{Discussion} 
\subsection{Shapes of the $Z_{\rm max}$ distributions and $\Delta$-scaling}

The behavior of the cumulants is of interest in the context of $\Delta$-scaling proposed for studying criticality
in finite systems \cite{d-p,d-e,d-n}.
Probability distributions $P(s_{\rm max})$ of the extensive order parameter $s_{\rm max}$ for different ``system sizes''
$\langle s_{\rm max}\rangle$ obey $\Delta$-scaling if they can be converted to a single scaling function
$\Phi(z_{(\Delta)})$ by the transformation
\begin{equation}
\langle s_{\rm max}\rangle^{\Delta}P(s_{\rm max})=\Phi(z_{(\Delta)})
\equiv \Phi \left( \frac{s_{\rm max}-\langle s_{\rm max}\rangle}
{\langle s_{\rm max}\rangle^{\Delta}} \right),
\end{equation}
where $1/2 \leq \Delta \leq 1$.

The $\Delta$-scaling method has been applied   
to distributions of the largest fragment charge
with expectations that the distributions obey the $\Delta=1/2$ scaling
in the ordered (low temperature) phase and the $\Delta=1$ scaling
in the disordered (high temperature) phase \cite{d-p,frankland,ma05,pichon06,gruyer13}.
The transition between the two scaling regimes would signal the presence of
a phase transition. The percolation model contradicts
such expectations \cite{jb}.
In the percolation disordered phase no $\Delta$-scaling is
observed for the largest cluster size.
Concerning the ordered phase, the $\Delta=1/2$ scaling
can only be observed for different system sizes at a fixed
value of the control parameter. This requirement
is difficult to realize experimentally.
Moreover, in systems of small sizes corresponding to
nuclear systems, this limiting scaling behavior is
violated as a consequence of surface effects.

\begin{figure*}[ht]
\includegraphics[width=8.0cm]{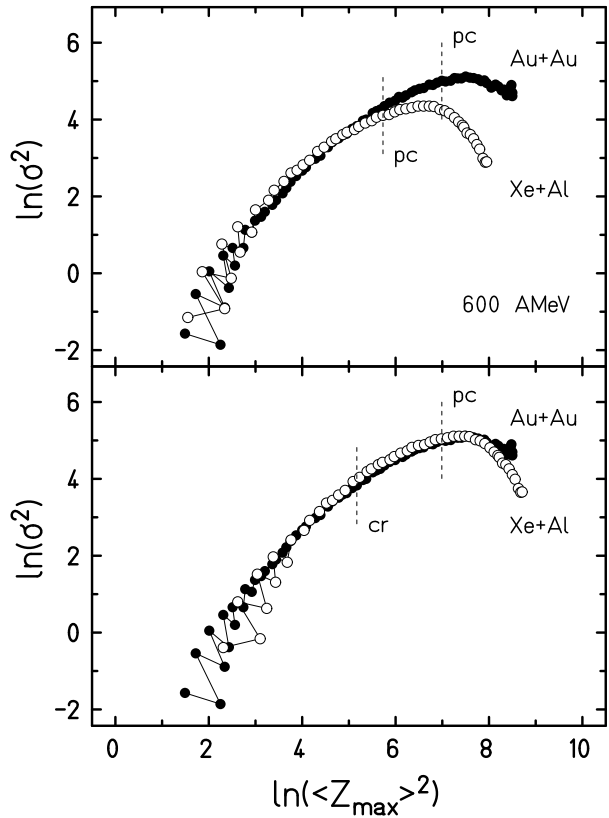}
\caption{Natural logarithm of the variance as a function of the
natural logarithm of the squared mean value of the largest atomic number
$Z_{\rm max}$ recorded in $^{197}$Au + $^{197}$Au (filled circles)
and $^{131}$Xe + $^{27}$Al collisions (open circles) at $600$ MeV/nucleon (top panel)
and after shifting of the $^{131}$Xe data with the value 0.76 in both dimensions (see text). The data
symbols represent the results for individual values of $Z_{\rm bound}$ in the range
from 4 to $Z_{\rm proj}$ + 1. The positions of the pseudocritical points are indicated for the two systems 
in the top panel, and the critical (cr) and pseudocritical (pc) points for $^{197}$Au + $^{197}$Au 
are given in the bottom panel (dashed vertical lines).}
\label{fig:9}
\end{figure*}

The present experimental data with events sorted 
according to $Z_{\rm bound}$ do not show $\Delta$-scaling
features in any $Z_{\rm bound}$ range. This can be concluded from Fig.~4, considering that
$K_{3}=\mathrm{const}$ and $K_{4}=\mathrm{const}$ are
necessary conditions for $\Delta$-scaling~\cite{jb}. Another condition required for
$\Delta$-scaling is a linear correlation of the natural logarithms of the variance and
squared mean value of the variable considered as an order parameter. This correlation 
is shown in Fig.~9 for the experimental results for $^{197}$Au + $^{197}$Au and 
$^{131}$Xe + $^{27}$Al collisions, both at $600$~MeV/nucleon and after sorting according
to $Z_{\rm bound}$. The positions of the pseudocritical points are indicated in the 
top panel of the figure.
For $^{197}$Au + $^{197}$Au, the corresponding $Z_{\rm bound}$ is directly taken from Fig.~4.
For $^{131}$Xe + $^{27}$Al, the pseudocritical $Z_{\rm bound}$ reported for $^{124}$Sn + Sn 
in Ref.~\cite{pietrzak} was used and a 8\% correction was applied corresponding to the
ratio of the atomic numbers of the Xe and Sn projectiles. 

The similarity of the two correlations is even better appreciated if the $^{131}$Xe + $^{27}$Al
result is scaled according to the atomic numbers $Z$ of the Au and Xe projectiles. Their 
ratio 79/54 that enters squared in the arguments of the logarithms leads to a linear shift by 0.76 in both, $x$ and $y$,
dimensions (Fig.~9, bottom panel). The similarity of the two functions reflects the invariance with 
respect to the projectile $Z$ that was observed for the fragment multiplicities and correlations in Ref.~\cite{sch}. 
The smooth variation of the slopes of the correlations
is very similar to the percolation result reported in Fig.~7 of Ref.~\cite{jb}. 
On the basis of the observed near perfect descriptions of the experimental data with percolation (cf. Figs.~6 and 8), 
even the remarkable quantitative agreement is not surprising. The location of the turnover at 
${\rm ln}(\langle Z_{\rm max} \rangle^2) \simeq 7$ and ${\rm ln}(\sigma^2) \simeq 5$ near the pseudocritical point 
is well reproduced. The observed experimental correlation is also similar to the 
results reported in Ref.~\cite{frankland} for the Xe+Sn reactions
at incident energies between 25 and 50 MeV/nucleon (Fig.~4 of Ref.~\cite{frankland}). With these reactions, 
after sorting according to the measured total transverse energy of light charged particles, an interval of approximately
$5.4 < {\rm ln}(\langle Z_{\rm max} \rangle^2) < 7.6$ was covered while, in the present case, the correlation extends as far down
as ${\rm ln}(\langle Z_{\rm max} \rangle^2) \simeq 2$. 
The definition of $Z_{\rm bound}$ causes the staggering that is observed there for the very small values of $Z_{\rm bound}$. 

\begin{figure*}[ht]
\includegraphics[width=11.0cm]{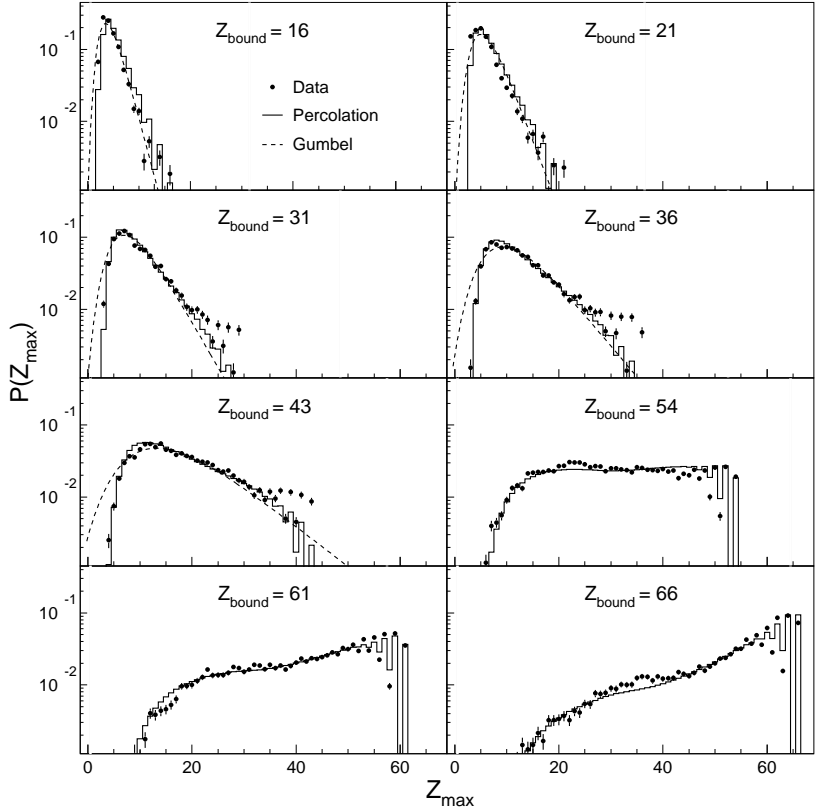}
\caption{Probability distributions $P(Z_{\rm max})$: comparison of percolation results (histograms) for the indicated values of $Z_{\rm bound}$
with the experimental data for all systems (filled circles) and with Gumbel distributions fitted to the experimental data 
for the cases $Z_{\rm bound} \le 43$ (dashed lines). 
}
\label{fig:10}
\end{figure*}

The trend towards positively skewed $Z_{\rm max}$ distributions expected in the disordered regime 
is approximately realized, both in the experimental data
and in the percolation model describing them. The asymptotic value $K_3 \simeq 1.6$
of the skewness for small $Z_{\rm bound}$ (Fig.~4) or large bond-breaking
probabilities $p_b$ (Fig.~1) is larger than the $K_3 = 1.14$ of the Gumbel distribution
in the continuous limit but of the same order of magnitude. 
The Gumbel distribution permits rather satisfactory descriptions of the experimental 
$Z_{\rm max}$ probability distributions
in this range of $Z_{\rm bound}$ as illustrated in Fig.~10 for selected cases. 
Only the sharp drop of the distribution at small $Z_{\rm max}$ cannot be reproduced.
The calculations were performed for
system sizes $Z_0 = 31,$~36, 48, 52, 57, 64, 70, and 73 for $Z_{\rm bound} = 16,$~21, 31, 36, 43, 54, 61, and 66, respectively.

In Ref.~\cite{frankland}, scaling has been further studied for the recorded most central collisions as a function
of the bombarding energy. For $^{197}$Au + $^{197}$Au at incident energies 40 to 80 MeV/nucleon, a linear scaling
with $\Delta=1$ has been observed (Fig.~13 in Ref.~\cite{frankland}). This trend is qualitatively continued by the
present data taken at 600 MeV/nucleon. If extended to ${\rm ln}(\langle Z_{\rm max} \rangle^2) \simeq 2$, the linear fit shown in that figure will
reach ${\rm ln}(\sigma^2) \simeq -1$, there coinciding with the most central bins of the present data set (Fig.~9).
Gaussian distributions are, however, not observed here.
At small $p_b$, neither the skewness nor the kurtosis excess do approach the 
vanishing values $K_3 = K_4 = 0$ characterizing the Gaussian distribution (Figs.~1 and 4).
The Gaussian distribution is reached as the asymptotic value with bond percolation only
in the continuous limit or for large systems with periodic boundary conditions~\cite{jb}.
The experimental situation is dominated by the transition to negatively skewed $Z_{\rm max}$
distributions (Fig.~10). At very large $Z_{\rm bound}$, corresponding to smaller excitation energies in  
the experiment, the partitioning probability decreases approximately exponentially with decreasing $Z_{\rm max}$.
This is well reproduced with percolation.

\subsection{Remarks on bimodality}

Bimodality in distributions of the largest fragment size
or other quantities expected to be closely correlated with
the order parameter is considered as a promising signature
of first-order phase transition~\cite{lopez,cho,gross}.
Experimental examinations of the largest-fragment charge distribution have, however, not ascertained
yet such a signal up to now. The presence of bimodality has been
reported for distributions of some other quantities as, e.g., the charge asymmetry
between the two or three largest fragments, and the asymmetry ratio between heavy and light fragments~\cite{bord,bell,ma05,pichon06,lopez}.
In the ALADIN data on $^{197}$Au + $^{197}$Au at $1000A$~MeV, a bimodal distribution of 
$Z_{\rm max}-Z_{2}-Z_{3}$ has been found in the transition region $Z_{\rm bound}=53-55$~\cite{lopez,tra}.
The bimodal behavior of this variable is also observed in bond percolation in which only a second-order phase transition 
is present. It has been identified as a finite-size effect obeying a power-law with the known 
value $\nu = 0.88$~\cite{sta} of the critical exponent describing the divergence of the correlation length~\cite{tra}. 

\begin{figure*}[ht]
\includegraphics[width=9.cm]{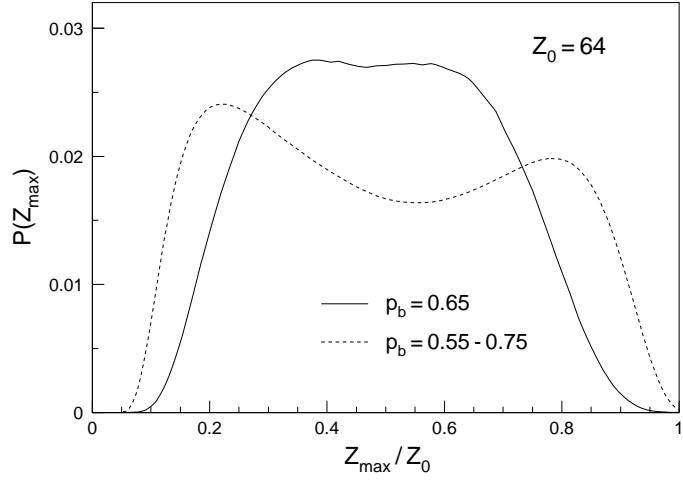}
\caption{Bond percolation for $Z_{0}=64$.
The probability distribution of $Z_{\rm max}$
at the transition point $p_{b}=0.65$ (solid line)
and for the bond probability distributed over a finite range
(dashed line).}
\label{fig:11}
\end{figure*}

As expected for the continuous percolation transition,
the distribution of $Z_{max}$ does not exhibit bimodality.
The shape of the transitional distribution is characterized
by a wide plateau as illustrated in Fig.~2(a) and Fig.~11.
Such characteristics are observed when events are sorted
according to the control parameter $p_{b}$.
If $p_{b}$ is dispersed in a sample of events,
the shape of $P(Z_{max})$ can be very different.
In particular, a bimodal shape may be observed as illustrated
in Fig.~11. It serves as a warning against using
wide bins for event sorting (see, e.g., Ref.~\cite{lef2009}) or sorting variables
that are not well correlated with the control parameter.

The latter applies to the study of projectile fragmentation in $^{197}$Au + $^{197}$Au collisions at 
energies between 60 and 100 MeV/nucleon that reported on bimodal behavior of the heaviest-fragment 
distributions~\cite{bonnet2009}. Since a strict canonical sampling is not possible in the experiment 
a scheme of selecting and weighting event groups has been applied with the aim to generate 
equivalent-to-canonical data samples. The obtained enhancements of the $Z_{max}$ distributions at values 
of 0.9 $Z_{0}$ and 0.3 $Z_{0}$ with excitation energies below 2 and above 8 MeV/nucleon, respectively, correspond to 
residue production in peripheral collisions and in highly fragmented processes at the threshold to vaporization. 
Without the strict sorting conditions that have been applied, these event groups will be 
characterized by rather different temperatures and system sizes. The reported observations, based on 
retaining only tails of their distributions, thus represent the studied reactions only very indirectly~\cite{mallik2018}.

\begin{figure*}[ht]
\includegraphics[width=15.cm]{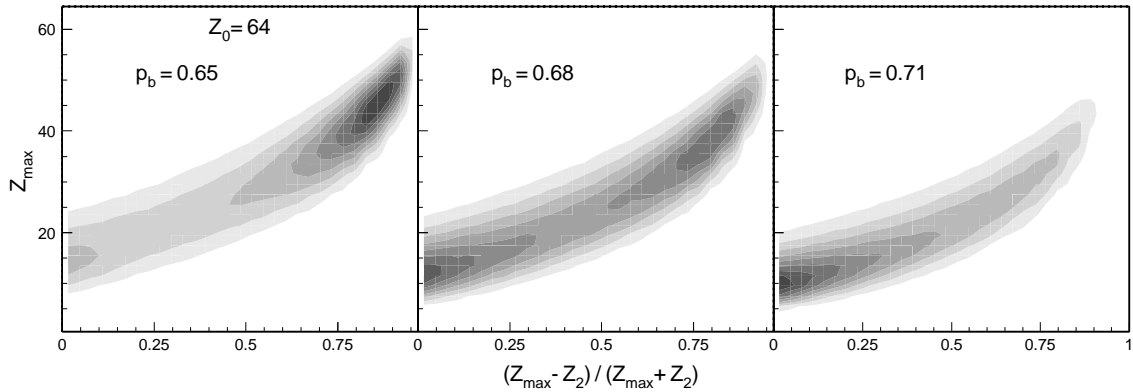}
\caption{Bond percolation for $Z_{0}=64$. Correlations between
$Z_{max}$ and the size asymmetry of the two largest
clusters in the transition region.}
\label{fig:12}
\end{figure*}

Even when percolation events are sorted by the control
parameter, various size asymmetry variables exhibit
bimodal behavior in the transition region.
As an example, Fig.~12 shows the correlation between $Z_{max}$
and the size asymmetry of the two largest fragments
$(Z_{max}-Z_{2})/(Z_{max}+Z_{2})$.
A bimodal structure of this distribution is clearly observed
at $p_{b}=0.68$.
It should be noted that the projection onto
the $Z_{max}$ axis does not reveal this bimodality.
A similar degree of bimodality is observed for a
much larger system with $16^{3}$ sites, suggesting
that this feature is not generated by finite size effects.
Correlations of this kind were examined experimentally for
fragmentation of projectiles in $^{197}$Au + $^{197}$Au and Xe + Sn reactions
at 80 MeV/nucleon with qualitatively similar results \cite{pichon06}.

Another example concerns the asymmetry between
the total sizes of large and small fragments.
Following the prescription applied to Xe + Sn central
collisions~\cite{bord}, fragments with $Z\geq13$ are considered as large
and fragments with $3\leq Z\leq12$ as small.
The evolution of their normalized difference distributions near the percolation
transition is illustrated in the left diagram of Fig.~13.
The system size of $100$ sites is comparable to the
total charge of the investigated nuclear system.
Also in this case a bimodal structure
is predicted by the percolation model.
The right diagram shows the qualitatively similar result
that is observed when the clusters with $Z=1-2$
are additionally included in the group of light fragments.

\begin{figure*}[ht]
\includegraphics[width=11.cm]{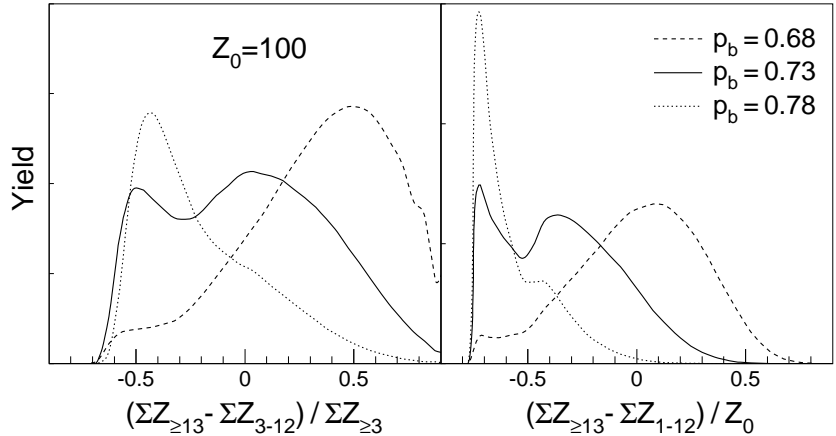}
\caption{Bond percolation for $Z_{0}=100$:
distributions of the normalized differences between the sum of atomic numbers of large fragments with $Z\geq13$ and 
the sum of atomic numbers of small fragments with $3\leq Z\leq12$ (left panel) or with $1\leq Z\leq12$ (right panel),
calculated for three values of the bond breaking probability $p_b$ below, near, and above the pseudocritical point.}
\label{fig:13}
\end{figure*}

The presented percolation simulations demonstrate that
bimodalities observed in distributions of the asymmetry variables are not necessarily associated
with a first-order phase transition. A similar conclusion was reached in Ref.~\cite{lef2008}
based on calculations with the quantum-molecular dynamics transport model for 
$^{197}$Au + $^{197}$Au collisions. There the authors concluded that fluctuations
introduced by elementary nucleon-nucleon collisions cause the
bifurcation observed in the distributions of fragment-charge asymmetries. Given the
good reproduction even of the higher-order distribution parameters, these effects are
realistically also included in the percolation description.

\subsection{Critical behavior in the coexistence zone}

The simultaneous appearance of signals expected for first- and second-order phase transitions in finite systems is known 
since long ago~\cite{mim,bug,das2002,gulm2005}. It has very recently been discussed again by investigating the 
lattice-gas model in addition to the percolation and thermodynamic models used here (cf. Figs.~1 and~2)~\cite{gupta2018}. 
All three models were shown to provide qualitative descriptions of  
experimental data while the transition points are indicative of either a first- or a second-order transition, 
depending on the model that is applied. 

A possible solution to the puzzle appeared when calculations with the lattice-gas model demonstrated the existence 
of critical-like regions within the coexistence zone of the phase diagram~\cite{mim,carmona,das2002,gulm2005,wieloch}. 
Scaling behavior was observed along a line that, in the case of small systems, extends from near the thermodynamical 
critical point of the phase diagram toward lower temperatures and lower densities into the coexistence region. 
It appears as an extension of the so-called Kert\'{e}sz line that is 
observed in lattice-gas and molecular-dynamics models at temperatures and densities above their critical 
points~\cite{kertesz,campi1997,campi2003}. 

According to the lattice-gas applications to fragmentation reactions, the associated 
critical temperatures are of the order of 5 MeV~\cite{wieloch} or 6 to 8 MeV~\cite{das2002}, i.e. far below the expected
critical temperature in a temperature-vs-density phase diagram, even for small systems 
(cf., e.g., Refs~\cite{jaqaman,schnack, furuta}). As noticed by Le Neindre {\it et al.}, these temperatures are comparable 
to the critical temperatures appearing in analyses based on Fisher scaling~\cite{leneindre2007}. In fact, the 
values reported in Refs.~\cite{kleine2002,elliott2002,elliott2003} are between 4.75 and about 8 MeV. 
Apparently, the "critical" disassemblies identified by searching for the scaling features of 
power-law type fragment spectra are located at or near the extension of the Kert\'{e}sz line into the 
coexistence zone. The observed broad range of reported temperatures is to be expected because different reaction 
types and experimental techniques 
will lead to different approaches of the critical line or region. There are, in addition, the uncertainties associated 
with determining temperatures from the observed fragment properties and yields. 

Also in the case studied here, the location of the line of critical-like behavior in a temperature vs density phase diagram is 
rather uncertain. However, the precision observed for the reproduction of fluctuation properties of percolation up to 
fourth order is remarkable.
Campi {\it et al.}, in their search for generic properties of the fragmentation of simple fluids, argue that the random 
breaking of bonds may be the simplest explanation for the appearance of percolation features in nuclear fragmentation~\cite{hyp}. 
The nucleon-nucleon collision dynamics introduces the required stochastic element in the present case of spectator fragmentation 
at relativistic energies.
In their later study~\cite{campi2003}, based on classical molecular dynamics calculations, the same authors have presented a 
scenario that involves the out-of-equilibrium expansion of the system, starting from an equilibrium configuration outside the 
coexistence zone. As the calculations show, the compositions generated at an early, high density stage are largely preserved in the final state.
Systems expanding from near the percolation critical region, the Kert\'{e}sz line in infinite systems, may thus appear with critical properties.
In an alternative scenario, the excited spectator matter equilibrates faster than it expands and cools, reaching the coexistence zone~\cite{schnack}.
Quantum-molecular-dynamics calculations that reproduce the experimental fragmentation patterns, may 
be capable of shedding more light on these possibilities.

\section{Conclusions}

The analysis of the ALADIN data on fragmentation processes of $^{197}$Au projectiles on 
heavy targets at energies between 600 and 1000 MeV/nucleon has been
focused on the fluctuations of the largest fragment size (charge or atomic number $Z$).
Cumulants of the largest fragment size distribution were examined
as a function of $Z_{\rm bound}$. Particularly valuable measures
are the higer-order cumulants,
skewness $K_{3}$ and kurtosis excess $K_{4}$.
The transitional distribution indicated by $K_{3}=0$ and a minimum of $K_{4}$
is characteristic of a phase transition. In percolation, it corresponds
to the pseudocritical point, and in a thermodynamic model, to the maximum
of the specific heat that is associated with a first-order transition.
Such a transition point is observed at $Z_{\rm bound}\simeq 54$ which, according to
Ref.~\cite{poch}, corresponds to excitation energies near 6 MeV/nucleon, associated with a temperature
between 5 and 7 MeV.

The cumulants $K_{3}$ and $K_{4}$ may be used as constraints for
comparisons with model predictions when system sizes
are not unequivocally determined.
They are shown to be not significantly affected by experimental conditions
and secondary decay effects.
Such constraints have been applied in the comparisons made with the bond
percolation model, intended to test the whole fragmentation pattern.

Fragment sizes and their event-to-event fluctuations observed in the experiment
are remarkably well reproduced by the bond percolation model.
The system sizes determined from the percolation analysis
are found to be in good agreement with experimental estimates.
The analysis suggests that near-critical events, corresponding to the true critical point, are located 
at $Z_{\rm bound}\simeq 36$. The associated excitation energy is about $10$~MeV/nucleon, leading to
a percolation critical temperature of again, approximately 6 to 7 MeV~\cite{poch,temp}. 

Owing to the high accuracy in describing the fragment-size properties,
the simple percolation model, containing a second-order phase transition, may still serve as a very useful reference model for studying
the phase behavior of fragmenting systems. In particular, it permits verifying the uniqueness of signatures proposed
for revealing the presence of a first-order phase transition. It needs to be stressed, however, that model comparisons, 
to be applicable for an experimental verification, should rely on event samples selected with measurable quantities.
These sorting quantities are inevitably dispersed over the control parameter of the model 
as, e.g., the temperature or the bond probability, which may significantly modify the expected signatures.

\begin{acknowledgments}
The authors would like to express their gratitude to the ALADIN Collaboration for the
permission to use the event-sorted fragmentation data of experiment S114.
This work was supported by the Polish Scientific Research Committee,
Grant No. 2P03B11023.
\end{acknowledgments}

\end{document}